\documentclass{Interspeech2024}
\usepackage{algorithm}
\usepackage{algpseudocode}
\usepackage{multirow}
\usepackage{booktabs}   
\usepackage{float}
\usepackage{cite}
\usepackage{tikz}
\usepackage{tkz-graph}
\usetikzlibrary{positioning}
\usetikzlibrary{bayesnet}
\usetikzlibrary{matrix}

\interspeechcameraready

\title{Phoneme Discretized Saliency Maps for Explainable Detection of AI-Generated Voice}

\name[affiliation={1,2}]{Shubham}{Gupta}
\name[affiliation={3,2,4}]{Mirco}{Ravanelli}
\name[affiliation={1,2,5}]{Pascal}{Germain}
\name[affiliation={1,2,3}]{Cem}{Subakan}

\address{
  $^1$Laval University, Canada, $^2$Mila-Québec AI Institute, Canada, \\ 
  $^3$Concordia University, Canada, $^4$Université de Montréal, Canada, $^5$Canada CIFAR AI Chair
  }
\email{shubham.gupta.1@ulaval.ca, mirco.ravanelli@concordia.ca,\\ \{pascal.germain, cem.subakan\}@ift.ulaval.ca%
\\
}


\begin{document}

\maketitle

\begin{abstract}
    
    %Time-frequency representations are very common input representations for audio classifiers. 
    
    %In this paper, we experimentally analyze the limitations of commonly used time-frequency representations for explainable detection of generated voice. Specifically, we show that when the artifacts are omni-present inside the time-frequency representation, the typical posthoc explanation methods such as gradient based saliency methods do not result in easy-to-understand explanations. We show with synthetic experiments that unless the artifacts are localized in time or frequency the produced explanations are very hard to understand for humans. We later showcase few alternative input representations which result in interpretations that result in more understandable model explanations. 
    In this paper, we propose Phoneme Discretized Saliency Maps (PDSM), a discretization algorithm for saliency maps that takes advantage of phoneme boundaries for explainable detection of AI-generated voice. We experimentally show with two different Text-to-Speech systems (i.e., Tacotron2 and Fastspeech2) that the proposed algorithm produces saliency maps that result in more faithful explanations compared to standard posthoc explanation methods. Moreover, by associating the saliency maps to the phoneme representations, this methodology generates explanations that tend to be more understandable than standard saliency maps on magnitude spectrograms.

    %Moreover, since we propose to associate the saliency maps to phoneme representations,     this methodology results in explanations that are more understandable compared to saliency maps on magnitude spectrograms. 
\end{abstract}

\section{Introduction}

% todo: consistency between saliency map and interpretation

In recent years, generative-AI has reached impressive levels of realism across a variety of domains \cite{cao2023comprehensive}. This includes text \cite{brown2020language}, conversational text \cite{chatgpt}, images \cite{DBLP:conf/cvpr/RombachBLEO22, DBLP:journals/cacm/GoodfellowPMXWO20}, and applications in the speech domain such as voice cloning / text-to-speech \cite{DBLP:conf/iclr/Chen0LLQZL21, DBLP:conf/interspeech/PaulSPS20, casanova2023yourtts}.

While these advanced generative modeling methods have relatively benign applications in areas such as education, movie production, or the video gaming industry, unfortunately, there also exists a plethora of malevolent use cases. These range from students cheating on homework using ChatGPT \cite{chatgptbot}, to fake news generation \cite{fakenews}, using deepfakes to instigate scandals \cite{scandals}, using AI-generated voice for scam calls \cite{scam}, and generating fake product reviews \cite{fakereview}. The perception in the general public against generative-AI has reached a point where Time magazine has ranked generative-AI as the third biggest global threat for 2023, and a significant threat to democracies around the world \cite{globalrisks} as they argue that generative-AI could be used to manipulate public perception in favor of malevolent actors.

The generative modeling techniques are slated to improve even further with the relentless pace of advancement in the machine learning literature, and therefore it is natural to expect that advances in generative-AI technology will only exacerbate social risks in the coming years. This is especially risky when human voices are replicated, as malevolent actors can use fake human voices for various malicious acts. 
\renewcommand{\thefootnote}{}
%\footnotetext{\textit{Code available at \href{https://github.com/shubham-gupta-30/pdsm}{https://github.com/shubham-gupta-30/pdsm}}}
\footnotetext{\textit{This research was enabled in part by support provided by the Digital Research Alliance of Canada (alliancecan.ca) and the Natural Sciences and Engineering Research Council of Canada (NSERC).}}

It is thus essential to develop countermeasures against systems that generate human voices. One such countermeasure is to use machine learning to automatically detect AI-generated voice. Detecting AI-generated voice with machine learning is a developing field. Existing works include  \cite{AlBadawy_2019_CVPR_Workshops}, which performs a bispectral analysis on the input audio to classify whether the input audio is generated by a TTS system or not. Another system in the literature performs classification by using weights of an ASR system \cite{wang2020deepsonar}. Also, ASVSpoof 2021 is a popular spoofed voice detection challenge, where the top-performing systems are GMM-based lightweight convolutional network, \cite{todisco2019asvspoof}, and rawnet2 \cite{DBLP:conf/icassp/Tak0TNEL21}. More recent works also include a spatio-temporal graph attention network \cite{DBLP:journals/corr/abs-2107-12710}, using pretrained models \cite{zhiqiang2022pre}, a self-distillation framework \cite{xue2023learning}, and a speaker verification model that aims to compare biometric characteristics~\cite{DBLP:journals/corr/abs-2209-14098}. 

Even though these methods give high accuracy in detecting AI-generated voice, their predictions are not interpretable or explainable, and therefore they remain black-box models that are opaque for end users. In this work, our focus is therefore particularly on faked voice detection, in a posthoc interpretable manner (i.e. producing interpretations for trained networks). A common way to process audio and speech signals is to use time-frequency representations such as the short-time-fourier-transform (STFT) or mel-frequency spectrograms. A straightforward approach for producing explanations is to apply standard posthoc interpretation methods such as Integrated Gradients \cite{sundararajan2017axiomatic}, GradientSHAP \cite{lundberg2017unified}, or Layerwise Relevance Propagation \cite{lrp} on a magnitude spectrogram. One such study has been conducted by \cite{Lim2022DetectingDV} for generated voices, however as we qualitatively and quantitatively show in this paper, producing heatmaps with standard posthoc interpretations typically results in interpretations that are difficult to understand.

As we discuss in the next section, the difficulty in producing interpretations for AI-generated voice stems from the fact that preconceived notions for what constitutes an acceptable explanation is absent. In this paper, we therefore propose \textbf{P}honeme \textbf{D}iscretized \textbf{S}aliency \textbf{M}aps (PDSM), which expresses the produced interpretations in terms of phonemes where the phoneme boundaries are extracted with an ASR model through a posterior distribution over phonemes for each time point \cite{DBLP:journals/corr/abs-2402-17735}. More specifically we show that, 

\begin{itemize}
    \item PDSM yields a more faithful explanation for the neural network decisions compared to standard saliency methods. In other words, PDSM explanations follow the neural network decisions more closely.
    \item PDSM expresses the neural network decision in terms of phonemes, and therefore produces more understandable explanations compared to standard saliency methods. 
\end{itemize}

We also release the dataset used in this paper for fostering future research on this topic.
\vspace{-.2cm}

%, unless the fakeness characteristics associated with the speech is localized on time and/or frequency. 

%What we however see is that alternative representations such as ... result in more understandable explanations. 

\section{Posthoc Interpretations for Audio}

\begin{figure}[t]
    \centering
    \resizebox{0.49\textwidth}{!}{
    \includegraphics[width=0.30\textwidth]{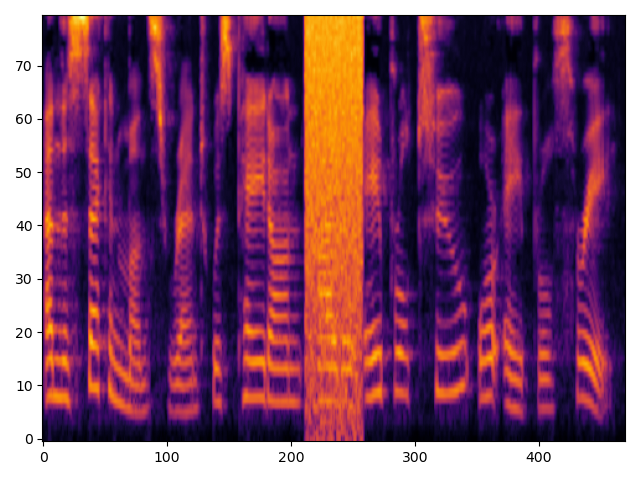}
    \begin{tikzpicture}
        \node (a) [] {\includegraphics[width=0.29\textwidth]{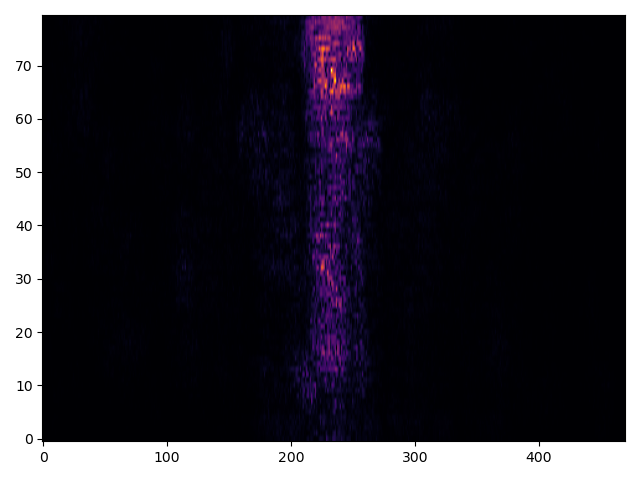}};
        \draw [color=white, -] (-.15,1.8) -- (-.15, -1.6);
        \draw [color=white, -] (.4,1.8) -- (.4, -1.6);
        \draw [color=white, <->] (-.15,-1) -- (.4, -1);
    \end{tikzpicture}
    }
    \caption{\textbf{(left)} Input Spectrogram, 
    \textbf{(right)} Interpretations obtained for a trained network with the GradSHAP algorithm. We overlay the boundaries of the time-limited Gaussian noise on top of the interpretations. 
    %\textbf{(right)} Interpretations obtained for an untrained network
    }
    \vspace{-.5cm}
    \label{fig:interpretations}
\end{figure}
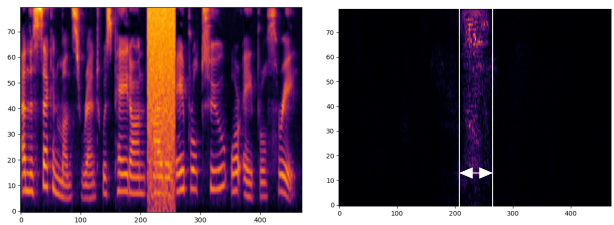

Magnitude spectrograms are a very common input domain representation for classifying audio signals. There exist several works in the literature that aim to produce interpretations on magnitude spectra \cite{l2i2022parekh, Mishra2017LocalIM, DBLP:journals/jfi/BeckerV0MLS24}. The goal is to identify time-frequency regions in the spectrogram that maximally contribute to the classifier decision. 

In image domain, the literature mainly deals with cases where the target object is clearly defined, and in such cases, the posthoc interpretation methods are able to produce interpretation results that are placed on the target object \cite{DBLP:journals/corr/SimonyanVZ13, ribeiro2016why, sundararajan2017axiomatic, Selvaraju_2019}. 

In the audio domain, we show an analogous example below, where the `target object' is clearly defined. We contaminate LJSpeech \cite{ljspeech17} utterances with time-limited standard Gaussian noise (this noise is added in the time domain, and the time boundaries are randomly sampled), and then train a model with CNN14 architecture to detect the presence of this noise. We then run a posthoc interpretation method for this trained classifier to highlight regions of the spectrogram which are deemed to be maximally important for the classifier. For this particular example, we used the GradSHAP algorithm \cite{lundberg2017unified}, but note that another posthoc interpretation algorithm could be used also to demonstrate the same thing. We show these results in Figure \ref{fig:interpretations}, where we observe that in this simple case the posthoc interpretation method clearly identifies the cause of the decision when applied on a trained network (right panel). 

%On the right panel we also show the interpretations obtained for an untrained network \cite{adebayo2020sanity}, and we see that the interpretation produced for the trained network is much more unequivocally focussed on the introduced Gaussian noise, compared to the untrained network. 

\subsection{Faithfulness of the interpretations}
To evaluate how much the interpretation reflects the salient causes that trigger the classifier's decision, the authors in \cite{l2i2022parekh} introduced a faithfulness metric. This metric is calculated by measuring the drop in class-specific output probabilities when the masked-out portion of the input is fed to the classifier. This amounts to calculating, 
\begin{align}
    \text{FF}_n:= f(X_n)_c - f(X_n\odot(1-M))_c, \label{eq:ff}  
\end{align}
where $f(.)_c$ denotes the classifier output for class $c$, and $M$ denotes the saliency map estimated by the interpretation method. If this metric is large, it signifies that the parts of the spectrogram highlighted by the interpretation method are highly influential for the classifier decision for class $c$, as the second term removes the parts of the input that are highlighted by the interpretation mask $M$. However, we notice that even when the saliency maps make intuitive sense, the faithfulness metric might have a very low value. For instance, we see from Figure \ref{fig:interpretations} that the produced interpretations point towards the part of the spectrogram that corresponds to the Gaussian noise which the classifier is trained to detect, and therefore, intuitively, this should result in large faithfulness value. However, the naive application of this metric typically results in poor faithfulness value because it fails to remove the Gaussian noise entirely, which the classifier is able to detect in the masked-out input (i.e. $X \odot (1-M))$. This can be potentially mitigated by discretizing the continuous saliency map to fill its bounding box shown in the right panel of Figure \ref{fig:interpretations}, thereby accurately encapsulating the target object. The core idea of the method proposed in this paper is to associate the high-energy regions in the saliency maps with time boundaries that accurately align with a notion of a `target object', namely phonemes. 

%(This aforementioned discretization constitutes a motivation for our proposed algorithm that we we describe in the next section) 

%\cem{We should show this on Figure 1 also, and explain in the caption.}

%We report the average faithfulness over all examples by reporting the average quantity $\text{FF}:= \sum_n \frac{1}{N} \text{FF}_n$. Larger is better. 

%We would like to also note that on slightly more complicated cases where the target objects display more variability compared to the case above but still correspond to the well-defined sound classes such as `dog barking', `siren', `baby cry' (e.g. classifying sound classes from the ESC50 dataset \cite{piczak2015dataset}) it is feasible to produce interpretations that are faithful \cite{l2i2022parekh, paissan2023posthoc}. 

As we show in the next section, in more complicated problems such as detection of AI-generated voice, where the `target object' is not clearly defined, evaluating the faithfulness of the interpretations and understanding the interpretations themselves prove to be difficult. We need techniques to be able to i) increase the understandability of  interpretations ii) measure the faithfulness accurately. For this purpose we propose to discretize the continuous saliency maps by using phoneme boundaries that encapsulate  high-energy portions of saliency maps.  We show that such discretization not only enhances understandability of interpretations, but also result in more faithful saliency maps.

\subsection{Producing Interpretations for AI-Generated Voice}
\begin{figure}[t]
    \centering
    \includegraphics[width=0.22\textwidth]{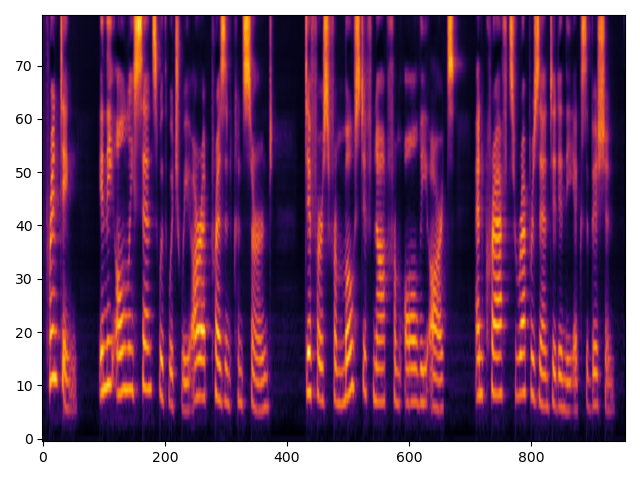}
    \includegraphics[width=0.22\textwidth]{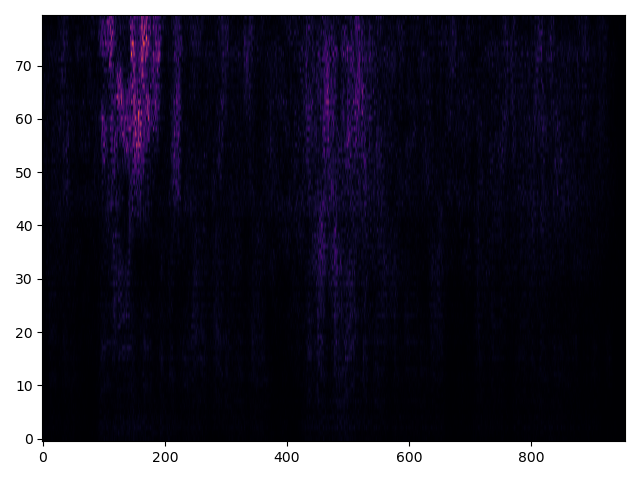}
    \caption{\textbf{(left)} Input Spectrogram for an AI-generated voice, \textbf{(right)} GradSHAP interpretations obtained for a CNN14 trained network}
    \vspace{-.4cm}
    \label{fig:interpretations-fake}
\end{figure}

We now consider producing posthoc interpretations for a classifier that detects AI-generated voice. For this purpose, we created a deepfake dataset using LJSpeech. We used $13,000$ utterances ($10,000$ train and $3,000$ test) from the LJSpeech Dataset as bonafide (real) speech. We generated $13,000$ fake utterances using the Tacotron2 model \cite{DBLP:conf/icassp/ShenPWSJYCZWRSA18} and another $13,000$ fake utterances using Fastspeech2 model \cite{DBLP:conf/iclr/0006H0QZZL21}), along with a HifiGAN vocoder. We used  a popular open-source speech package - SpeechBrain \cite{ravanelli2021speechbrain} for this purpose. We trained two CNN14 models \cite{DBLP:journals/taslp/KongCIWWP20} to distinguish between real and generated speeches, one for Tacotron2 and another for Fastspeech2. Both classifiers obtains a detection accuracy very close to $100 \%$ on the test set.  

%\cem{We need to specify what is the obtained accuracy on the test set}

In Figure \ref{fig:interpretations-fake}, we show GradSHAP interpretations obtained for an AI-generated voice correctly classified by our classifier. 
%We observe that the interpretations produced for the trained network are highly focused compared to the network. 
We observe that the interpretations produced for the sample (right panel) are unlike the time-localized Gaussian noise example, where the `target object' is clearly defined, so in this case, it is hard to ascertain i) the exact meaning of the produced interpretations, ii) if the produced interpretations are indeed faithful. We now propose Phoneme Discretized Saliency Maps and show how they help us alleviate both these issues. 

\section{Phoneme Discretized Saliency Maps (PDSM)}

In this section, we propose PDSM, a method that discretizes saliency map interpretations produced by attribution methods such as Integrated Gradients (IG), GradSHAP, and others, using phoneme posteriorgram (PPG) representations. 
%We show that the binary masks obtained from these discretized maps are not only more faithful than the original saliency maps but they are also more understandable. 

The phoneme posteriorgrams \cite{DBLP:journals/corr/abs-2402-17735} provide a posterior distribution over the spoken phoneme for each time frame in a spectrogram audio representation. PPGs can be obtained via an ASR model that is trained with phoneme labels (e.g. on the TIMIT dataset \cite{timit}) or can be obtained through forced alignment \cite{mcauliffe17_interspeech}. In this paper, we use an ASR model that is trained on TIMIT, which has a Phoneme Error Rate (PER) of $24.6\%$ averaged over TIMIT, Arctic and CommonVoice datasets  \cite{DBLP:journals/corr/abs-2402-17735}. 

% Given an utterance, we obtain mel spectrogram and PPG representations for this utterance. While we use the mel spectrogram as input to the classifier and After obtaining the saliency map and the phoneme posteriorgrams, we then associate the two. Intuitively, we would like to note that the the time axes for the saliency map and the posteriorgram match, so what remains to be done is to activate a phoneme boundary if there is non-negligible energy in the corresponding time region of the saliency map.
Given a saliency map $M$, obtained via a saliency method (e.g., IG) for a classifier decision on a speech utterance, we discretize $M$ by pooling its weights contained within phoneme boundaries. These phoneme boundaries are derived from the PPG representation of the utterance. We describe the exact procedure in Algorithm \ref{algo:pdsm} (note that we use the Python array notation). That is, $B[:, s_i:e_i]$ takes the entire first dimension of a matrix $B$, and takes the slice that runs between $s_i$ and $e_i$ for the second dimension. We also pictorially explain the algorithm in Figure \ref{fig:pdsm}. We would like to note that there are a few hyperparameters of this algorithm that affect the faithfulness of generated binary maps - specifically the choice of the \textit{preprocess} and the \textit{pool} functions. 

Once we obtaine a discretized Saliency map, we convert it into a binary mask by keeping only the $k$ most important phonemes as judged by the map.

\begin{algorithm}
\caption{Phoneme Discretization for Saliency Maps}
\label{algo:pdsm}
    \begin{algorithmic}
    \State \textbf{Input:} Saliency Map $M$ of shape $[F, T]$, PPG $X^p$ of shape $[N, T]$, number of phonemes to keep $k$.
    \State \textbf{Output:} Phoneme Discretized Saliency Map $\widehat M$.
    \State \textbf{Hyperparameters:}  functions $preprocess$, $pool$. \\
    \State $\bullet$ $\widetilde M = $ \textit{preprocess}$(M)$
    \State $\bullet$ Get time frames to phoneme alignment:\\
    \phantom.\hfill$\widetilde X^p_t = \arg \max_{i} X^p_{i,t}, \text{ for } t \in \{1, \dots, T\}\,.$ \hfill\phantom.
    \State $\bullet$ Get phonemes and their start and end boundaries:\\
    \hfill\phantom.$(p_1, s_1, e_1), \dots ,(p_n, s_n, e_n)$ from $(\widetilde X^p_{t})$\,.\hfill\phantom.
    \State $\bullet$ Let energy of phoneme $p_i$ be  $E(p_i)$ := \textit{pool}$(\widetilde M[:,\  s_i:e_i])$.
    \State $\bullet$ Let $J \subset \{1, 2, \dots n\}$, $|J| = k$ be the indices of $k$ highest energy phonemes.
    \State $\bullet$ Initialize $\widehat M$ = \textit{zeros}$(F, T)$\,.
    \State $\bullet$ $\widehat M[:, \ s_j:e_j] = 1$, $\forall j \in J$ 
    \State $\bullet$ Return $\widehat M$
    %\While{$t\geq T$}{
    %asd
    %}
    \end{algorithmic}
\end{algorithm}

Note that $N$ is the vocabulary size of the PPG representation, and $n$ is the number of phonemes in this utterance. We first preprocess the saliency map - which may involve thresholding the map to remove small values and/or passing the map through an $abs$ (absolute value) function. We noticed that these exact preprocessing decisions are crucial and serve as hyperparameters for our algorithm. For our experiments with fake voice generated using Tacotron2, we found that thresholding the maps and not taking their $abs$ resulted in more faithful discrete maps. However, for FastSpeech2, we obtained better maps by thresholding and taking an $abs$ of the original saliency maps.

Similarly, the choice of the pooling function affects the faithfulness of the obtained discrete maps. We found mean pooling to be more effective for Tacotron2, but sum pooling worked better for FastSpeech2.

\tikzstyle{block} = [draw, fill=lightgray, rectangle, 
    minimum height=3em, minimum width=4em]
\tikzstyle{sumt}   = [circle, minimum width=8pt, draw, inner sep=0pt, path picture={\draw (path picture bounding box.east) -- (path picture bounding box.west) (path picture bounding box.south) -- (path picture bounding box.north);}]
\begin{figure}[h!]
    \centering
    \begin{tikzpicture}
        \node [block, fill=white] (ppg)  {\includegraphics[width=0.11\textwidth]{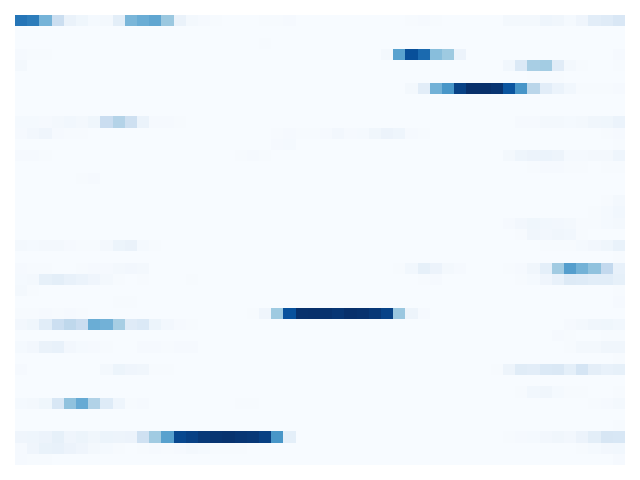}};
        \node () [below of=ppg, xshift=-.5cm] {$X^p$};
        \node [block, fill=white, right of=ppg, xshift=2cm] (sm) { \includegraphics[width=0.13\textwidth]{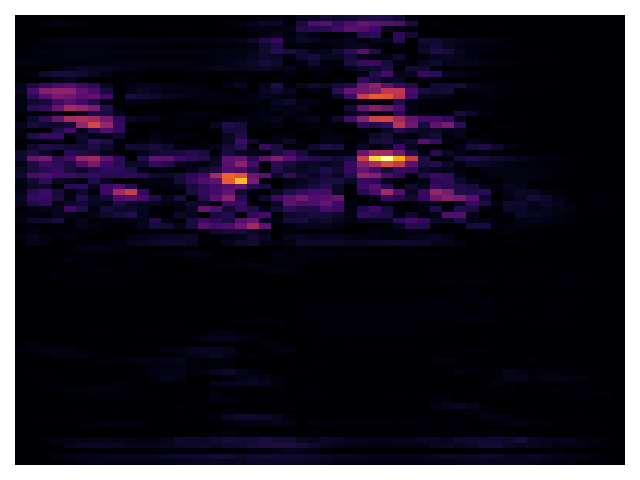} };
        \node () [below of=sm, xshift=-.5cm, yshift=-.1cm] {$M$};
        \node [block, below of=sm, yshift=-1cm] (pprocess) {preprocess(.)};
        \node [block, fill=white, below of=pprocess, yshift=-1cm] (sm2) {\includegraphics[width=0.11\textwidth]{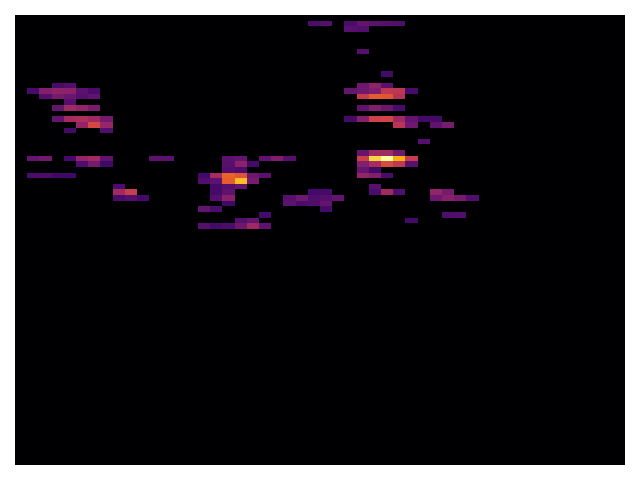}};
        \node () [above of=sm2, xshift=-.5cm, yshift=+.1cm] {$\widetilde M$};
        
        \node [block, fill=white, right of=sm2, xshift=-4.0cm, yshift=-0cm] (pbs) { \includegraphics[width=0.11\textwidth]{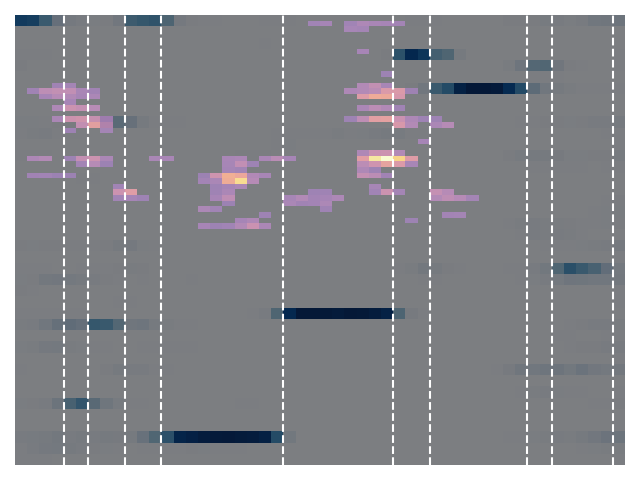} };
        \node [block, below of=pbs, yshift=-1cm, xshift=-0cm] (pool) {pool(.)};
        \node [block, fill=white, right of=pool, yshift=0cm, xshift=1.2cm] (dmask) { \includegraphics[width=0.11\textwidth]{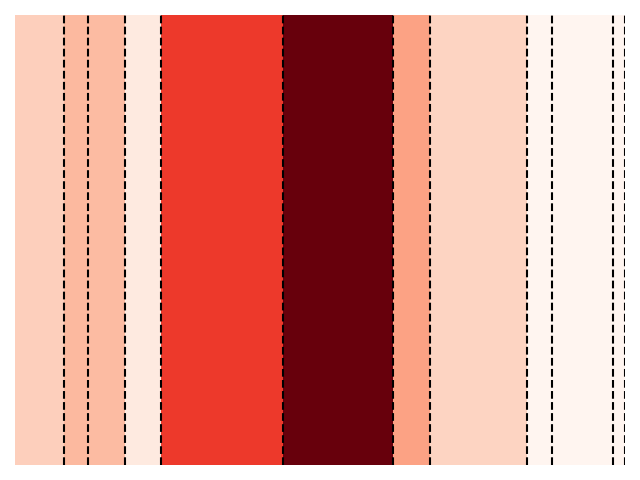} };
        \node [block, fill=white, right of=dmask, yshift=0cm, xshift=1.7cm] (bmaskm) { \includegraphics[width=0.11\textwidth]{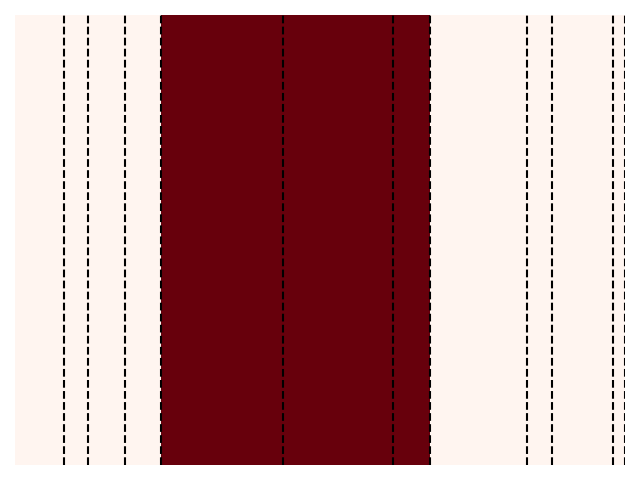} };
        \node () [above of=bmaskm, xshift=-.5cm, yshift=+.1cm] {$\widehat M$};

        \draw [->] (sm) -- (pprocess); 
        \draw [->] (pprocess) -- (sm2); 
        \draw [->] (sm2) -- (pbs);
        \draw [->] (ppg) -- (pbs);
        \draw [->] (pbs) -- (pool);
        \draw [->] (pool) -- (dmask);
        \draw [->] (dmask) -- node [anchor=west, rotate=90] {\footnotesize{Binarize}} (bmaskm);
    \end{tikzpicture}
    \caption{Steps of the PDSM algorithm.}
    \vspace{-.5cm}
    \label{fig:pdsm}
\end{figure}
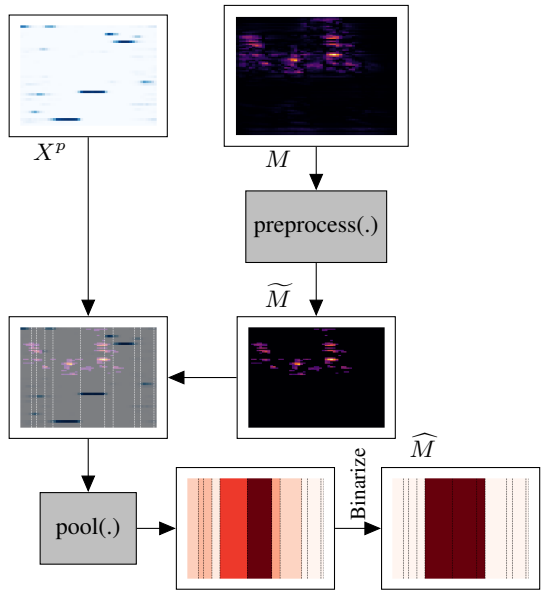

%TODO
\begin{figure*}[h!]
    \centering
    \includegraphics[width=0.24\textwidth]{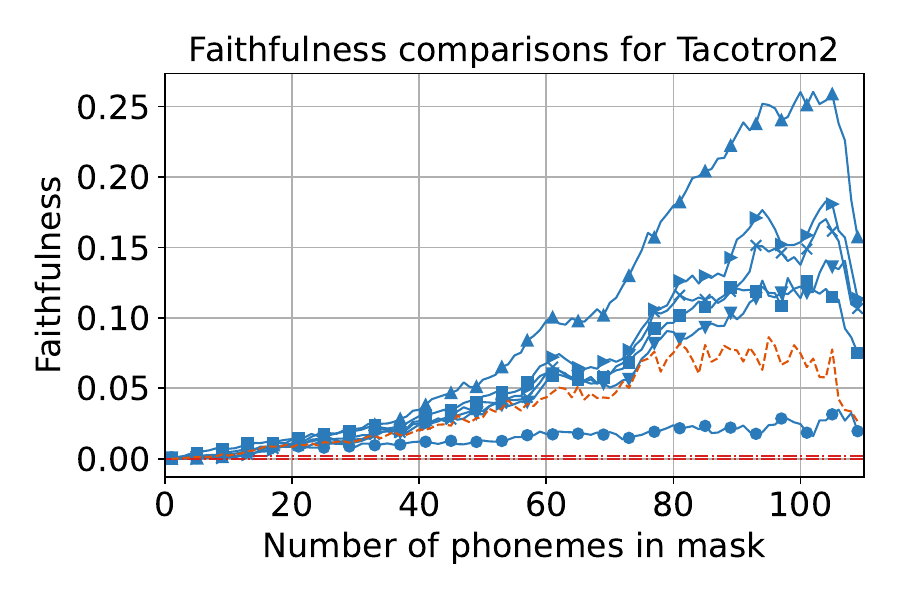}
    \includegraphics[width=0.24\textwidth]{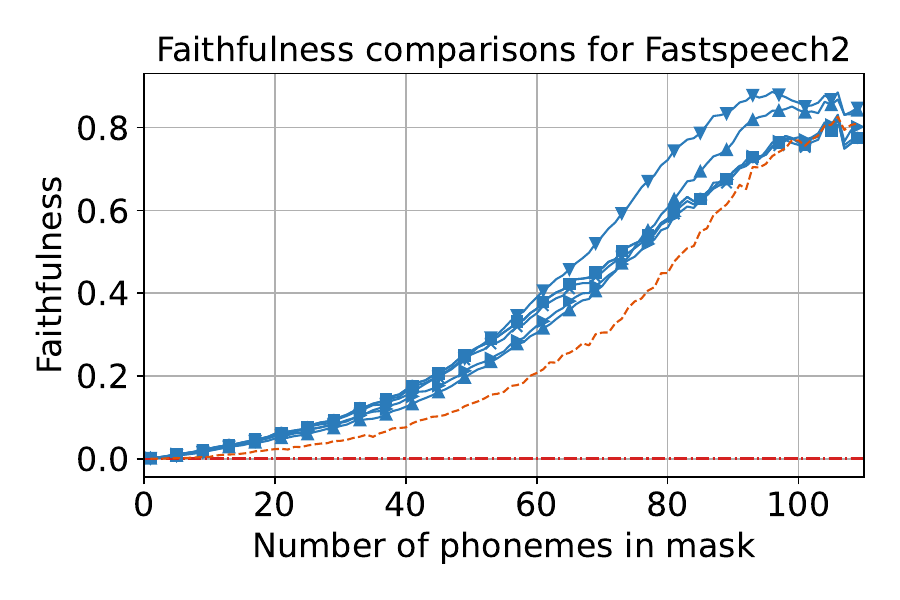}
    \includegraphics[width=0.24\textwidth]{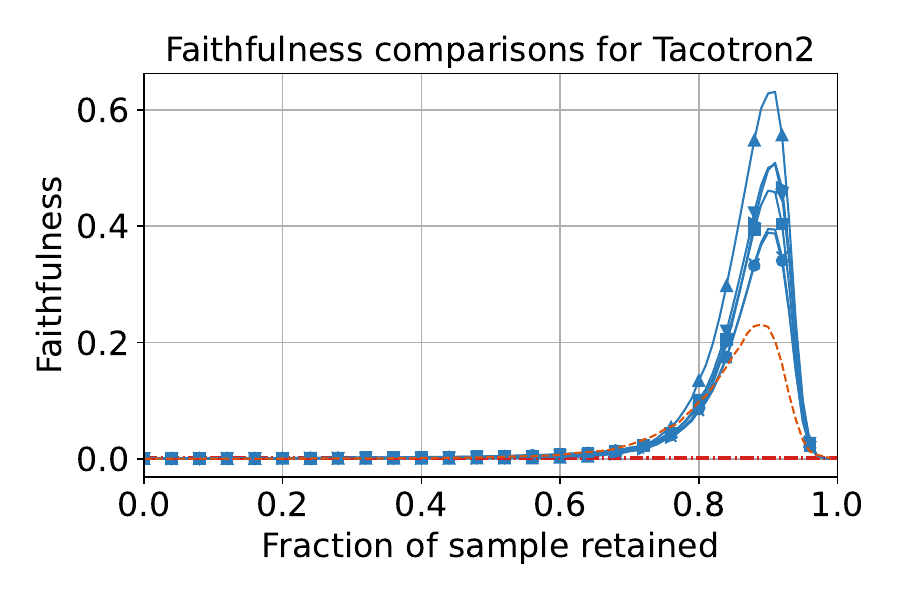}
    \includegraphics[width=0.24\textwidth]{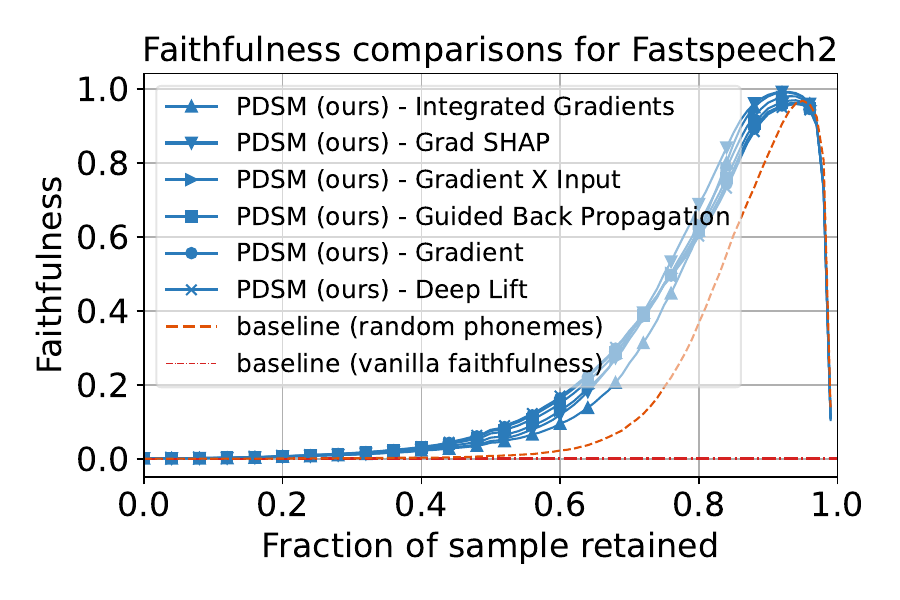}
    \vspace{-.3cm}
    \caption{\textbf{(left-most, left-center)} Analysis of faithfulness with respect to different numbers of phonemes retained in the binary mask for Tacotron2 and FastSpeech2 respectively. \textbf{(right-center, right)} Same analysis of faithfulness normalized over the length of audio sample, thus representing fraction of audio retained for the binary mask.   
    }
    \label{fig:FFvsnph}
\end{figure*}

%\section{Alternative Representations}
\section{Evaluating the Faithfulness of PDSMs}
\label{sec:faithfulness}

To systematically measure the faithfulness obtained on the real/fake classification problem, we calculate the faithfulness metric defined in Equation \eqref{eq:ff} for standard saliency methods and our proposed phoneme discretized saliency maps (PDSM). To also assess the effectiveness of picking phonemes for the binary mask through PDSM we compare our methods with the faithfulness obtained by randomly assigning phonemes to the mask as a baseline. We compared faithfulness for various saliency methods (IG \cite{sundararajan2017axiomatic}, GradSHAP \cite{lundberg2017unified}, GradInput \cite{molnar2022}, GuidedBP \cite{DBLP:journals/corr/SpringenbergDBR14}, Gradient Saliency \cite{molnar2022}, DeepLift \cite{DBLP:conf/icml/ShrikumarGK17}). In Figure \ref{fig:FFvsnph}, we compare the behavior of the faithfulness metric for different number of phonemes retained (shown on the x-axis of the plots) in the discretized map. We observe in the left-most and the middle-left panel that PDSM obtains better faithfulness metrics for all values of $k$ (number of phonemes in the mask) compared to the baseline that randomly picks $k$ phonemes (except the vanilla Gradient saliency method). In the middle-right and right-most panels of Figure \ref{fig:FFvsnph}, we also do this same comparison, but normalized by the length of the utterance. We again see that PDSM results in much more faithful explanations as compared to faithfulness obtained from randomly picking phonemes or faithfulness resulting from original salilency maps.

In Table \ref{tab:FFnumbers}, we compare the faithfulness of PDSM compared to the straightforward application of different saliency methods vs. their application after going through the PDSM algorithm. We observe that the application of the PDSM algorithm significantly increases the faithfulness of the saliency map compared to direct computation of faithfulness on the original saliency maps. The disparity between TT2 and FS2 can potentially be explained by the differences in underlying classifier confidence.

%\cem{Here we should specify more once we have the numbers} 

% \begin{table*}[t]
% \caption{Comparison of average faithfulness of PDSM vs. alternatives}
% \label{tab:FFnumbers}
% \centering
% %\resizebox{13.9cm}{!}{
% \begin{tabular}{l|c|c|c||c|c|c|}
%  &  \multicolumn{3}{|c|}{Tacotron2} & \multicolumn{3}{|c|}{FastSpeech2} 
%  \\ \hline \hline
% \textbf{Interpretation Method} & \textbf{Plain} & \textbf{Random Phonemes} &  \textbf{PDSM} & \textbf{Plain} & \textbf{Random Phonemes} &  \textbf{PDSM}  \\
% \hline \hline 
% IG  &  & &0.10 & - & - & -  -  -   \\ \hline 
% GradSHAP & & &0.052  & - & - & - -  -  \\ \hline  
% GradInput & & & 0.066 & - & - & - -  -  \\ \hline  
% GuidedBackProp & &  &0.056 & - & - & - -  -  \\ \hline
% Saliency & & &0.015 & - & - & - -  -  \\ \hline
% Deep Lift & & &0.059 & - & - & - -  -  \\ \hline
% 
% \end{tabular}
% 
% 
% %\end{table}
% \end{table*}

\begin{table*}[t]
\caption{Comparison of average faithfulness of PDSM (in bold) vs. plain saliency methods (in parantheses), for Tacotron 2 (denoted as TT2), and Fastspeech 2 (denoted as FS2). On the right most column we also include the random phoneme selection baseline (denoted with RP) discussed in Section \ref{sec:faithfulness}. }
\vspace{-.3cm}
\label{tab:FFnumbers}
\centering
\resizebox{0.99\textwidth}{!}{
\begin{tabular}{l|c|c|c|c|c|c|c}
% &  \multicolumn{3}{|c|}{Tacotron2} & \multicolumn{3}{|c|}{FastSpeech2} 
 %\\ \hline \hline
& IG & GradShap &  GradInput & GuidedBP & Gradient & DeepLift &  RP    \\
\hline \hline 
\textbf{TT2} & \textbf{0.10} \; ($10^{-5}$)  & \textbf{0.052} \; ($2\cdot 10^{-3}$) &\textbf{0.066} \; ($-1.9\cdot 10^{-5}$) & \textbf{0.056} \; ($-2.4\cdot 10^{-5}$) & \textbf{0.015} \; ($-2.4\cdot 10^{-5}$) &\textbf{0.059} \; ($-2.4\cdot 10^{-5}$) & 0.037 \\ \hline 
\textbf{FS2} & \textbf{0.35}  \; ($10^{-3}$) &\textbf{0.396} \; ($5\cdot 10^{-4}$) & \textbf{0.337} \; ($6\cdot 10^{-4}$) & \textbf{0.352} \; ($2\cdot 10^{-4}$) & \textbf{0.352} \; ($2\cdot 10^{-4}$) & \textbf{0.346} \; ($2.8\cdot 10^{-5}$)  & 0.27  \\ \hline 

\end{tabular}
}
%\end{table}
\end{table*}

% cem: here we should comment on the results

%cem : maybe a subsection? \subsection{Analyzing the faithfulness behavior}

\section{Evaluating the Understandability of PDSMs}

\begin{figure}[h!]
    \centering
    \includegraphics[width=.45\textwidth]{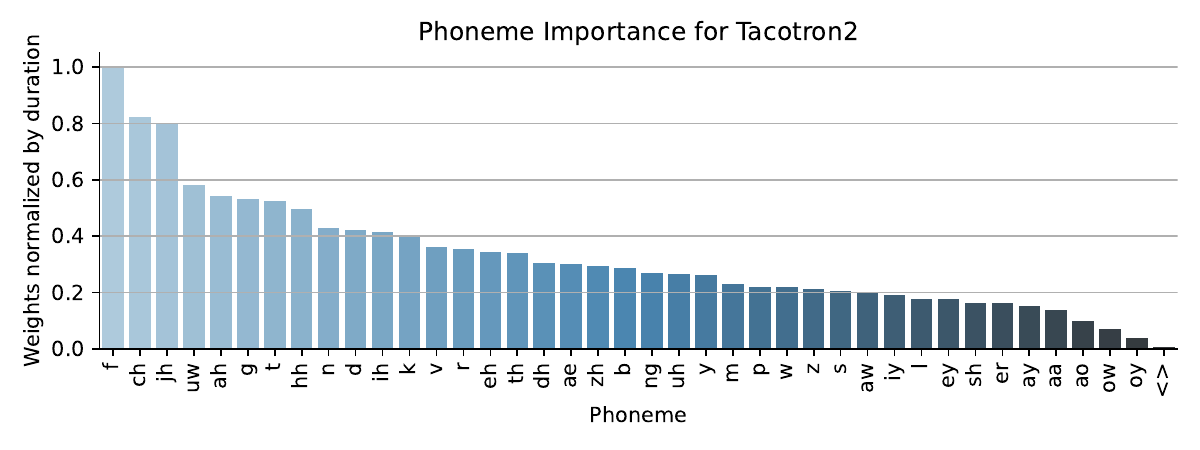}
    \includegraphics[width=.45\textwidth]{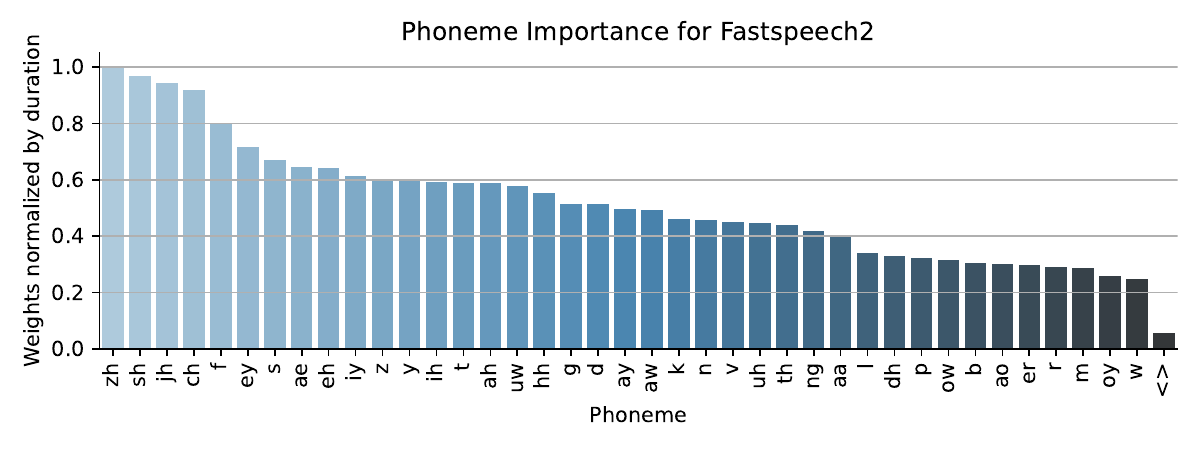}
    \vspace{-.2cm}
    \caption{Global Phoneme Importances Obtained by Tacotron2 and FastSpeech2. Note that $<>$ indicates silence. }
    \vspace{-.4cm}
    \label{fig:globalunderstandability}
\end{figure}

First, we show that PDSMs provide understandable explanations for AI-generated voice detection. In Figure \ref{fig:globalunderstandability}, we provide a population-level analysis to show the highest-ranked phonemes in terms of saliency weights for Tacotron2 and FastSpeech2. To obtain these plots, we compute the total pooled energy assigned to a phoneme by the saliency maps obtained using Integrated Gradients accumulated over the entire test set. We then normalize this total energy by the duration (measured in number of time frames occupied in mel representation) of this phoneme computed over the entire test set. We notice two trends that ascertain the fact that PDSM is giving phoneme importance as expected. i) We observe that in both cases fricatives are assigned large relative importance, which, in our personal experience, are typically problematic in TTS systems. ii) We see that a smaller number of phonemes are assigned high relative importance in Tacotron2 as compared to FastSpeech2 which makes sense as Tacotron2 is known to sound more realistic. 
\vspace{-.2cm}

\begin{figure}[h!]
    \centering
    \includegraphics[width=.45\textwidth]{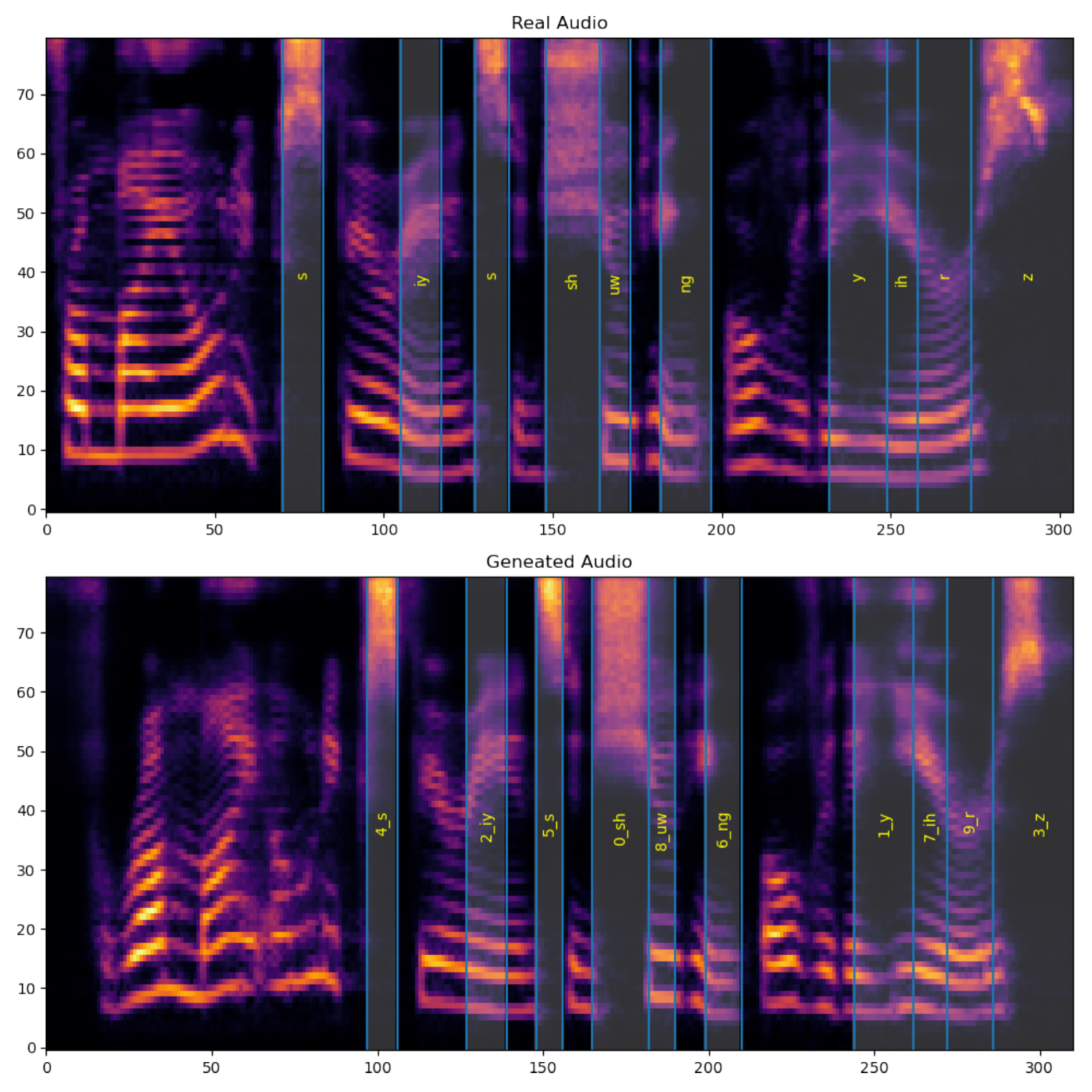}
    \vspace{-.0cm}
    \caption{Sample wise analysis of understandability of PDSM interpretations. \textbf{(top)} Mel-Spectrogram for a real audio \textbf{(bottom)} Mel-Spectrogram for a generated audio that corresponds to the same utterance. In both figures we overlay the phoneme boundaries for the top 10 most important phonemes ranked using PDSM. We show the corresponding rank in the ordering (from 0 to 9) followed by the phoneme identity. We see that the algorithm accurately focuses on phonemes which appear unnatural on the spectrogram, for example, fricatives like /s/,  and others like /iy/}
    \vspace{-.7cm}
    \label{fig:sampleunderstandability}
\end{figure}

Finally, we show an analysis on a particular example for FastSpeech2. In Figure \ref{fig:sampleunderstandability}, we show the phonemes with largest assigned saliency values. We observe that when contrasted with the same phoneme boundaries in the real speech, PDSM accurately detects phonemes that appear unnatural. We also observe that the algorithm points out to phonemes which have been given large global importance as shown in the bottom panel of Figure \ref{fig:globalunderstandability}.  

\section{Conclusions and Discussions}
In this paper, we proposed PDSM, a method that discretizes saliency maps through phoneme boundaries to produce faithful and understandable explanations for complex classification tasks where the users do not have preconceived notions to be able to understand the saliency maps. We showed that PDSM produces more faithful results compared to standard posthoc interpretation methods. We  also showed that the generated interpretations can be broken down in terms of phonemes which can help users understand the neural network decision much more easily. 

Even though in this paper we have focused on using explainable detection of AI-generated voice, PDSM is a general algorithm that can potentially be used in other complex speech/audio classification tasks. Therefore as future work, we would like to expand the scope of PDSM to cover complex speech classification tasks such as emotion recognition, and speech-related medical applications.

\bibliographystyle{IEEEtran}
\bibliography{mybib}

\end{document}